\documentclass[twocolumn,amsmath,amssymb]{snp}
\usepackage{graphicx}
\usepackage{dcolumn}
\usepackage{bm}
\begin{document}

\title{{\Large Isospin effects in the disappearance of flow as a function of colliding geometry}}


\author{\large Sakshi Gautam$^1$}
\author{\large Aman D. Sood$^2$}
\author{\large Rajeev K. Puri$^1$}
\email{rkpuri@pu.ac.in}
\affiliation{$^1$Department of Physics, Panjab University,
Chandigarh - 160014, INDIA} \affiliation{$^2$SUBATECH, Laboratoire de Physique Subatomique et des Technologies Associ\'{e}es, Universit\'{e} de Nantes - IN2P3/CNRS - EMN \\
4 rue Alfred Kastler, F-44072 Nantes, France} \maketitle

\section*{Introduction}
With the availability of radioactive ion beam (RIB) facilities,
one has possibility to study the properties of nuclear matter
under the extreme conditions of isospin asymmetry. Heavy-ion
collisions induced by the neutron rich matter provide a unique
opportunity to explore the isospin dependence of in-medium nuclear
interactions, since isospin degree of freedom plays an important
role in heavy-ion collisions through both nuclear matter equation
of state (EOS) and in-medium nucleon-nucleon (nn) cross-section.
After about three decades of intensive efforts in both nuclear
experiments and theoretical calculations, the equation of state of
isospin symmetric matter is now relatively well determined. The
effect of isospin degree of freedom on the collective transverse
in-plane flow as well as on its disappearance \cite{krof89} (there
exists a particular incident energy called \emph{balance energy}
(E$_{bal}$) or \emph{energy of vanishing flow} (EVF) at which
transverse in-plane flow disappears) has been reported in the
literature \cite{pak97,gaum10}, where it was found
 that neutron-rich systems have higher  E$_{bal}$ compared to neutron-deficient
  systems at all colliding geometries varying from central to peripheral
ones. The effect of isospin degree of freedom on  E$_{bal}$ was
found to be much more pronounced at peripheral colliding
geometries
 compared to central ones. Since colliding geometry has a significant role in the isospin effects so here we aim to understand
  the isospin effects in E$_{bal}$ as well as on its mass dependence over
  full range of colliding geometry varying from central to
 peripheral ones. The present study is carried out within the
 framework of isospin-dependent quantum molecular dynamics (IQMD)
 model \cite{hart98}.


\section*{Results and discussion}
\begin{figure}[!t]
\centering \vskip 0cm
\includegraphics[angle=0,width=7cm]{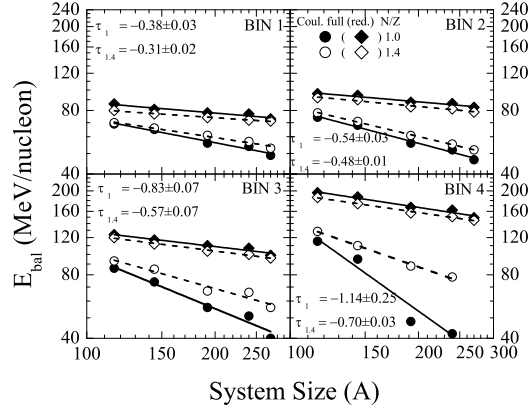}
\vskip 0cm \caption{E$_{bal}$ as a function of combined mass at
different impact parameter bins.}\label{fig1}
\end{figure}
  We have simulated the reactions of
$^{24}$Mg+$^{24}$Mg, $^{58}$Cu+$^{58}$Cu, $^{72}$Kr+$^{72}$Kr,
$^{96}$Cd+$^{96}$Cd, $^{120}$Nd+$^{120}$Nd, $^{135}$Ho+$^{135}$Ho,
having N/Z = 1.0 and reactions $^{24}$Ne+$^{24}$Ne,
$^{58}$Cr+$^{58}$Cr, $^{72}$Zn+$^{72}$Zn, $^{96}$Zr+$^{96}$Zr,
$^{120}$Sn+$^{120}$Sn, and $^{135}$Ba+$^{135}$Ba, having N/Z =
1.4, respectively. The colliding geometry is divided into four
impact parameter bins of 0.15 $<$ $\hat{b}$ $<$ 0.25 (BIN 1),
 0.35 $<$ $\hat{b}$ $< $0.45 (BIN 2), 0.55 $<$ $\hat{b}$ $<$ 0.65 (BIN 3),
and 0.75 $<$ $\hat{b}$ $<$ 0.85 (BIN 4), where $\hat{b}$ =
b/b$_{max}$. Figure 1 displays the mass dependence of E$_{bal}$
for these impact parameter bins. The solid (open) circles indicate
 E$_{bal}$ for systems with lower (higher) neutron content.
  E$_{bal}$ follows a power law behavior
$\varpropto$ A$^{\tau}$ for both N/Z = 1 and 1.4 at all colliding
geometries. Isospin effects are clearly visible as neutron-rich
system has higher E$_{bal}$ throughout the mass range. The
magnitude of isospin effects increases with increase in the impact
parameter and mass of the system for a given impact
   parameter bin. One also sees that the difference between
$\tau_{1.0}$ and $\tau_{1.4}$ increases with increase in the
impact parameter. The solid (open) diamonds represent E$_{bal}$
calculated with reduced Coulomb calculations for systems with
lower (higher) neutron content. Lines are power law fit
$\varpropto$ A$^{\tau}$. The values of $\tau_{1.0}$ ($\tau_{1.4}$)
are -0.17$\pm$ 0.02 (-0.14$\pm$ 0.01), -0.17$\pm$ 0.02 (-0.19$\pm$
0.02), -0.24$\pm$ 0.03 (-0.26$\pm$ 0.01), and -0.31$\pm$ 0.03
(-0.29$\pm$ 0.02) for BIN 1, BIN 2, BIN 3, and BIN 4,
respectively. Interestingly, we find that the magnitude of isospin
effects is now nearly same throughout the mass range and also
throughout the range of colliding geometry. We also see that the
enhancement in the E$_{bal}$ (by reducing Coulomb) is more at
higher impact parameter compared to lower one for a given mass and
also the enhancement in the E$_{bal}$  is more in heavier systems
as compared to lighter systems for a given bin. Moreover,
throughout the mass range and range of colliding geometry, the
neutron-rich systems
 have less E$_{bal}$ as compared to neutron-deficient systems when we reduce the Coulomb \cite{gaum210}. This trend is quite the opposite to the one
 which we have when we have full Coulomb. This clearly shows that
 the enhancement in the isospin effects at peripheral colliding
 geometries is due to the Coulomb potential.
  \begin{figure}[!t]
\centering \vskip 0cm
\includegraphics[angle=0,width=7cm]{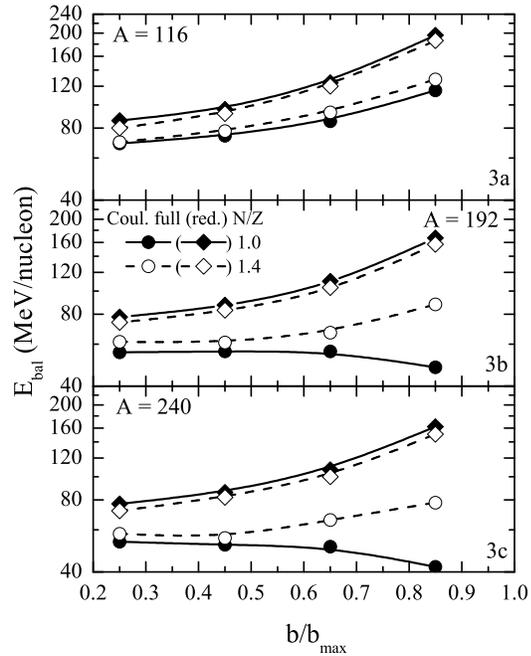}
\vskip 0cm \caption{E$_{bal}$ as a function of impact parameter
for different colliding masses.}\label{fig1}
\end{figure}

 In figs. 2a, 2b, and 2c, we display E$_{bal}$ as a function
of $\hat{b}$ for masses 116, 192, and 240, respectively, for both
full and reduced Coulomb. Symbols have the same meaning as in the
fig. 1. Form the figure, we find that:(i) for a given mass (eg.
A=116),
 the difference between E$_{bal}$ for systems with different N/Z (in case of reduced Coulomb) remains almost constant
throughout the range of colliding geometries which indicates that
the effect of symmetry energy is uniform throughout the range of
$\hat{b}$. (ii) Comparing figs. 2a, 2b, and 2c, one finds that for
a given $\hat{b}$, the difference between E$_{bal}$ for systems
having different N/Z remains constant throughout the mass range
also which indicates that the effect of symmetry energy is uniform
throughout the mass range as well.
\section*{Acknowledgments}
 This work is supported by Indo-French project
no. 4104-1.

\end{document}